\documentclass[runningheads]{llncs}
\usepackage[T1]{fontenc}
\usepackage{hyperref}
\usepackage{color}

\usepackage{enumitem}
\usepackage{amsmath,amssymb}
\usepackage{mathtools}
\usepackage[algoruled,vlined,english,linesnumbered]{algorithm2e}
\usepackage{sagetex}
\setlist{leftmargin=*} 

\usepackage{tikz}
\usetikzlibrary{arrows,matrix}

\SetArgSty{textup}
\SetKwInput{Input}{Input}
\SetKwInput{Output}{Output}
\SetKwIF{If}{ElseIf}{Else}{if}{:}{else if}{else:}{end}
\SetKwFor{For}{for}{\string:}{end}

\newcommand{\ZZ}{\mathbb{Z}}
\newcommand{\QQ}{\mathbb{Q}}
\newcommand{\RR}{\mathbb{R}}
\newcommand{\CC}{\mathbb{C}}

\def\<#1>{\langle#1\rangle}
\newcommand{\C}{\mathcal{C}}
\DeclareMathOperator{\CNF}{CNF}
\DeclareMathOperator{\Id}{I}

\DeclareMathOperator{\Ob}{Ob}
\DeclareMathOperator{\Mor}{Mor}
\DeclareMathOperator{\ObS}{\mathbf{Ob}}

\begin{document}
\title{
How to automatise proofs of operator statements: Moore-Penrose inverse -- a case study
}
\titlerunning{How to automatise proofs of operator statements}
%
\author{Klara Bernauer\inst{1} \and
Clemens Hofstadler\inst{2,}\thanks{The second author was supported by the Austrian Science Fund (FWF): P32301.} \and
Georg Regensburger\inst{2}}
\authorrunning{K. Bernauer, C. Hofstadler, and G. Regensburger}
%
\institute{
     Johannes Kepler University Linz, Austria\\
     \email{klarahbernauer@gmail.com} 
\and
    Institute of Mathematics, University of Kassel, Germany\\
    \email{\{clemens.hofstadler,regensburger\}@mathematik.uni-kassel.de}
}
\maketitle
\begin{abstract}
We describe a recently developed algebraic framework for proving first-order statements about linear operators by computations with noncommutative polynomials.
Furthermore,  we present our new \textsc{SageMath} package \texttt{operator\char`_gb}, which offers functionality for automatising such computations.
We aim to provide a practical understanding of our approach and the software through examples, 
while also explaining the completeness of the method in the sense that it allows to find algebraic proofs for every true first-order operator statement.
We illustrate the capability of the framework in combination with our software by a case study on statements about the Moore-Penrose inverse, including classical facts and recent results, presented in an online notebook.

\end{abstract}
\section{Introduction}

In its section on the Moore-Penrose inverse, the Handbook of Linear Algebra~\cite[Sec.~I.5.7]{hogben2013handbook} lists, besides the 
defining identities of the Moore-Penrose inverse \eqref{eq:moore-penrose}, a number of classical facts:

\begin{enumerate}
\item Every $A \in \CC^{m \times n}$ has a unique pseudo-inverse $A^\dagger$.
\item If $A \in \RR^{m \times n}$, then $A^\dag$ is real.
\item If $A \in \CC^{m \times n}$ [\dots] has a full rank decomposition $A = BC$ [\dots], then $A^\dag$ can be evaluated using $A^\dag = C^* (B^* A C^*)^{-1} B^*$.
\item If $A \in \CC^{m \times n}$ [\dots] has an SVD $A= U\Sigma V^*$, then its pseudo-inverse is $A^\dag = V\Sigma^\dag U^*$ [\dots].
\item[\vdots]
\end{enumerate}

Now, imagine the following task:
Prove as many of these facts as possible, using only the defining identities and no additional references.
Trying to do this by hand is a non-trivial task for any non-expert.
However, a recently developed framework~\cite{RRH21,hofstadler2022universal} allows to reduce proving such statements to computations with noncommutative polynomials, which can be fully automated, using for example 
our newly developed software package \texttt{operator\char`_gb}.
These polynomial computations yield algebraic proofs that are not only valid for matrices but for any setting where the statement can be formulated (e.g., linear operators on Hilbert spaces, homomorphisms of modules, $C^*$-algebras, rings, etc.).
Based on this, starting only from the defining equations, the first author was able to prove a majority of the facts on the Moore-Penrose inverse in the Handbook in a fully automated way in form of her Bachelor's thesis~\cite{klara-thesis}.
Various examples of automated proofs of matrix and operator identities based on computations with noncommutative polynomials are also given in~\cite{SL20,schmitz-master}.


In this paper, we describe this approach of using computer algebra to automatise the process of proving first-order statements of matrices or linear operators like the ones listed above.
The goal of the paper is twofold.
Firstly, readers will be able to gain a practical understanding of the framework by reading Section~\ref{sec:moore-penrose-uniqueness} and~\ref{sec:moore-penrose-existence}, 
learning how to translate operator statements into polynomial computations and how to use the software to compute algebraic proofs.
In Section~\ref{sec:moore-penrose-uniqueness}, we also discuss previous work on using noncommutative polynomials for proving operator identities.
Secondly, we explain that the framework allows to prove every \emph{universally true} first-order operator statement using the semi-decision Procedure~\ref{proc:procedure}, showing that the approach is complete in this sense.

In particular, in Section~\ref{sec:logical-framework},
we give a self-contained description of the framework developed in~\cite{hofstadler2022universal}, with a particular focus on applicability.
We focus on the simple, yet in practice most common, case of so-called \emph{$\forall \exists$-statements}, which drastically reduces the complexity of the presentation compared to~\cite{hofstadler2022universal} where  arbitrary first-order formulas are treated.
We note that this is in fact no restriction as any first-order formula can be transformed into an equivalent $\forall \exists$-formula using the concept of Herbrandisation and Ackermann's reduction, as detailed in~\cite[Sec.~2.2,~2.3]{hofstadler2022universal}.

We also present our new \textsc{SageMath} package \texttt{operator\char`_gb}\footnote{available at \url{https://github.com/ClemensHofstadler/operator_gb}}, which provides functionality for Gröbner basis computations in the free algebra, similar to~\cite{letterplace}.
In addition, our package offers dedicated methods for automatising the proofs of operator statements.
Most importantly, it provides methods for finding elements of specific form in polynomial ideals~\cite{heuristics}.
This not only allows to automatically prove existential statements, but also to effectively model many properties of operators (e.g., conditions on ranges and kernels, injectivity and surjectivity, cancellability properties, etc.).
For further details, we refer to Section~\ref{sec:common-properties}, and to  Appendix~\ref{appendix} for the corresponding commands.

Finally, in Section~\ref{sec:case-study}, we illustrate the capability of the framework in combination with our software in form of a case study on statements regarding the Moore-Penrose inverse.
We have successfully automated the proofs of a variety of theorems, ranging from the classical facts in the Handbook~\cite[Sec.~I.5.7]{hogben2013handbook} over important characterisations of the reverse order law for the Moore-Penrose inverse~\cite{DD10} to very recent improvements of Hartwig’s triple reverse order law~\cite{cvetkovic2021algebraic} that were found with the help of our software.
We have assembled a Jupyter notebook containing the automated proofs of all statements, which is available at
\url{https://cocalc.com/georeg/Moore-Penrose-case-study/notebook}.


\section{From operator identities to noncommutative polynomials}
\label{sec:moore-penrose-uniqueness}

In~1920, E.H.~Moore~\cite{moore1920reciprocal} generalised the notion of the inverse of a matrix
from nonsingular square matrices to all, also rectangular, matrices.
This \emph{generalised inverse}, by Moore also called ``general reciprocal'', was later rediscovered by Roger Penrose~\cite{Pen55}, leading to the now commonly used name \emph{Moore-Penrose inverse}.

Moore established, among other main properties, existence and uniqueness of his generalised inverse and justified its application to linear equations.
However, Moore's work was mostly overlooked during his lifetime due to his peculiar and complicated notations, which made his results inaccessible for all but very dedicated readers.
In contrast, Penrose characterised this generalised inverse by four simple identities, yielding the following definition:
The \emph{Moore-Penrose inverse} of a complex matrix $A$ is the unique matrix $B$ satisfying the four \emph{Penrose identities}
\begin{align}\label{eq:moore-penrose}
    A B A = A, && B A B = B, && B^* A^* = AB, && A^* B^* = B A,
\end{align}
where $P^*$ denotes the Hermitian adjoint of a complex matrix $P$.
Typically, the Moore-Penrose inverse of $A$ is denoted by $A^\dagger$.

Using the Penrose identities, and their adjoint versions that follow, makes basic computations involving the Moore-Penrose inverse very simple.
For example, uniqueness can be showed as follows.
If $B$ and $C$ both satisfy~\eqref{eq:moore-penrose}, then
\begin{align}\label{eq:proof-uniqueness}
\begin{aligned}
  B &= BAB = BACAB = BACB^*A^* = BC^*A^*B^*A^* \\
    &= BC^*A^* = BAC = A^*B^*C = A^*C^*A^*B^*C \\
    &= A^*C^*BAC = CABAC = CAC = C.
\end{aligned}
\end{align}


At the end of the last century, people realised that matrix identities, or more generally identities of linear operators, can be modelled by noncommutative polynomials, and that computations like~\eqref{eq:proof-uniqueness} can be automated using algebraic computations involving such polynomials.
For example, in the pioneering work~\cite{helton1994rules,helton1998computer} polynomial techniques were used to simplify matrix identities in linear systems theory, and in~\cite{HS99} similar methods allowed to discover operator identities and to solve matrix equations.

Noncommutative polynomials are elements in a free (associative) algebra $R\<X>$ with coefficients in a commutative ring $R$ with unity and noncommutative indeterminates in a (typically finite) set $X$.
Monomials are given by words over $X$, that is elements in the free monoid $\<X>$, and multiplication is given by concatenation of words.
In particular, indeterminates still commute with coefficients, but not with each other.

Intuitively, a matrix or operator identity $B = C$, or equivalently $B - C = 0$, can be identified with the polynomial $f(b,c) = b - c$.
More generally, identities of composite operators can be translated into noncommutative polynomials by introducing a noncommutative indeterminate for each basic non-zero operator, and by uniformly replacing each operator by the respective indeterminate in the difference of the left and right hand side of each identity.
Potentially present zero operators are simply replaced by the zero in $R\<X>$.

For example, to express the Penrose identities~\eqref{eq:moore-penrose}, we introduce indeterminates $a, b, a^*, b^*$ to represent the matrices $A, B$, and their adjoints, and form the polynomials
\begin{align}\label{eq:moore-penrose-polynomials}
  &aba - a, && bab - b, && b^* a^* - ab, && a^* b^* - ba.
\end{align}

With this, the computation~\eqref{eq:proof-uniqueness} corresponds to the polynomial statement
\begin{align}\label{eq:proof-uniqueness-polynomials}
\begin{aligned}
  b - c =\;&-(bab - b) \;-\; b(aca - a)b \;-\; bac(b^* a^* - ab) \;-\; b(c^*a^* - ac)b^*a^* \\
  -\;&bc^*(a^*b^*a^* - a^*) \;+\; b(c^*a^* - ac) \;-\; (a^*b^* - ba)c  \;-\; (a^*c^*a^* - a^*)b^*c\\
 +\;& a^*c^*(a^*b^* - ba)c \;+\; (a^*c^* - ca)bac \;+\; c(aba - a)c \;+\; (cac - c).
\end{aligned}
\end{align}
This shows that $b-c$ can be represented as a two-sided linear combination of the polynomials encoding that $B$ and $C$ satisfy the Penrose identities for $A$.

\begin{remark}\label{remark:involution}
To model the involution $*$ on the polynomial level, we introduce an additional indeterminate for the adjoint of each basic operator and simplify all operator expressions using the following identities before translating them into polynomials.
\begin{align}\label{eq:involution}
  (P+Q)^\ast = P^* + Q^*, && (PQ)^\ast = Q^\ast P^\ast, && (P^\ast)^\ast = P.
\end{align}
Furthermore, whenever an identity $P = Q$ holds, then so does the adjoint identity $P^* = Q^*$, and 
these additional identities have to be translated into polynomials as well.
Thus, to express that $B$ is the Moore-Penrose inverse of $A$ on the polynomial level, we have to add to~\eqref{eq:moore-penrose-polynomials} the polynomials corresponding to the adjoint identities.
Since the last two Penrose identities are self-adjoint, this yields the two additional elements $a^* b^* a^* - a^*$ and $b^* a^* b^* - b^*$.
We note that these additional polynomials can be essential for proofs and also appear in~\eqref{eq:proof-uniqueness-polynomials}.
\end{remark}

Algebraically, the relation~\eqref{eq:proof-uniqueness-polynomials} means that the polynomial $b-c$ lies in the (two-sided) ideal generated by the polynomials encoding that $B$ and $C$ are Moore-Penrose inverses of $A$.
We call such a representation of an ideal element in terms of the ideal's generators a \emph{cofactor representation}.

It is always the case, that, if an operator identity follows from given identities by arithmetic operations with operators (i.e., addition, composition, and scaling), then the polynomial corresponding to this identity is contained in the ideal.
However, not all polynomials that lie in the ideal correspond to valid operator identities, because, in contrast to computations with actual operators, computations with polynomials are not restricted and all sums and products can be formed. Obviously, elements of the ideal that do not comply with the formats of the matrices (or, more generally, the domains and codomains of the operators) cannot correspond to identities of operators.
Thus, \emph{a priori}, when proving an operator identity by verifying \emph{ideal membership} like in~\eqref{eq:proof-uniqueness-polynomials}, one has to ensure that every term appearing in a cofactor representation respects the restrictions imposed by the operators.

Algorithmically, ideal membership of commutative polynomials can be decided by Buchberger’s algorithm~\cite{Buc65} computing a Gröbner basis of the ideal.
In contrast, ideal membership of noncommutative polynomials is only semi-decidable in general.
This is a consequence of the undecidability of the word problem.
More precisely, \emph{verifying} ideal membership of noncommutative polynomials is always possible, using a noncommutative analog of Buchberger’s algorithm~\cite{Mor85,MZ98} to enumerate a (possibly infinite) Gröbner basis.
However, \emph{disproving} ideal membership is generally not possible.
Nevertheless, if a polynomial can be verified to lie in an ideal, then, as a byproduct, a cofactor representation of the polynomial in terms of the generators can be obtained.
This representation serves as a certificate for the ideal membership and can be checked independently.

Our \textsc{SageMath} software package \texttt{operator\char`_gb} allows to certify ideal membership of noncommutative polynomials by computing cofactor representations.
We illustrate its usage to compute the representation given in~\eqref{eq:proof-uniqueness-polynomials}.
To generate the polynomials encoding the Penrose identities, the package provides the command \texttt{pinv}.
Furthermore, it allows to automatically add to a set of polynomials the corresponding adjoint elements, using the command \texttt{add{\char`\_}adj}.
\begin{sageverbatim}
# load the package
sage: from operator_gb import *

# create free algebra
sage: F.<a, b, c, a_adj, b_adj, c_adj> = FreeAlgebra(QQ)

# generate Moore-Penrose equations for b and c
sage: Pinv_b = pinv(a, b, a_adj, b_adj)
sage: Pinv_c = pinv(a, c, a_adj, c_adj)
# add the corresponding adjoint statements
sage: assumptions = add_adj(Pinv_b + Pinv_c)

# form a noncommutative ideal
sage: I = NCIdeal(assumptions)

# verify ideal membership of the claim
sage: proof = I.ideal_membership(b-c)

# print the found cofactor representation
sage: pretty_print_proof(proof, assumptions)
\end{sageverbatim}
{{%
\abovedisplayskip=0pt plus 3pt 
\abovedisplayshortskip=0pt plus 3pt 
\belowdisplayskip=0pt plus 3pt 
\belowdisplayshortskip=0pt plus 3pt 
\begin{align*}
\text{\texttt{b{ }{-}{ }c{ }={ }}}
&\text{\texttt{({-}c{ }+{ }c*a*c){ }+{ }b*c{\char`\_}adj*({-}a{\char`\_}adj{ }+{ }a{\char`\_}adj*b{\char`\_}adj*a{\char`\_}adj)}}\\
&\text{\texttt{{-}{ }b*a*c*({-}a*b{ }+{ }b{\char`\_}adj*a{\char`\_}adj){ }{-}{ }b*({-}a{ }+{ }a*c*a)*b{ }}}\\
&\text{\texttt{+{ }b*({-}a*c{ }+{ }c{\char`\_}adj*a{\char`\_}adj){ }{-}{ }b*({-}a*c{ }+{ }c{\char`\_}adj*a{\char`\_}adj)*b{\char`\_}adj*a{\char`\_}adj}}\\
&\text{\texttt{{-}{ }({-}b{ }+{ }b*a*b){ }+{ }({-}c*a{ }+{ }a{\char`\_}adj*c{\char`\_}adj)*b*a*c}}\\
&\text{\texttt{{-}{ }({-}a{\char`\_}adj{ }+{ }a{\char`\_}adj*c{\char`\_}adj*a{\char`\_}adj)*b{\char`\_}adj*c{ }+{ }c*({-}a{ }+{ }a*b*a)*c}}\\
&\text{\texttt{{-}{ }({-}b*a{ }+{ }a{\char`\_}adj*b{\char`\_}adj)*c{ }+{ }a{\char`\_}adj*c{\char`\_}adj*({-}b*a{ }+{ }a{\char`\_}adj*b{\char`\_}adj)*c}}
\end{align*}}}

\begin{remark}
The computed representation is equal to~\eqref{eq:proof-uniqueness-polynomials} up to a reordering of summands.
\end{remark}

The correctness of a computed cofactor representation can be verified easily by expanding it, which only requires basic polynomial arithmetic.
Our package allows to do this using the command \texttt{expand\_cofactors}.

\begin{sageverbatim}
# reusing the assumptions and proof from above
sage: expand_cofactors(proof, assumptions)
\end{sageverbatim}
{{%
\abovedisplayskip=0pt plus 3pt 
\abovedisplayshortskip=0pt plus 3pt 
\begin{align*}
\text{\texttt{b - c}}
\end{align*}}}

The pioneering work mentioned above exploited the fact that the operations used in the noncommutative version of Buchberger’s algorithm respect the restrictions imposed by domains and codomains of operators, cf.~\cite[Thm.~25]{helton1998computer} or~\cite[Thm.~1]{SL20}. 
Thus, using Buchberger's algorithm, proving an operator identity can be reduced to verifying ideal membership of the corresponding polynomial.

Only recently it was observed that in fact any verification of ideal membership, even one that does not comply with the domains and codomains of the operators, allows to deduce a correct statement about linear operators, provided that all initial polynomials correspond to actual operator identities~\cite{RRH21}.
This implies that the verification of the ideal membership can be done completely independently of the operator context.

In particular, this also means that the cofactor representation given in~\eqref{eq:proof-uniqueness-polynomials} immediately yields the uniqueness statement of the Moore-Penrose inverse of a complex matrix.
Moreover, since the polynomial computation is independent of the concrete operator context, this representation also proves a corresponding statement in every setting where it can be formulated.
For example, we immediately obtain an analogous result for bounded linear operators between Hilbert spaces or for elements in $C^*$-algebras.
In fact, the most general setting that is covered by the polynomial computation is that of \emph{morphisms} in a \emph{preadditive semicategory}.

\begin{definition}
A \emph{semicategory} $\C$ (also called \emph{semigroupoid}) consists of
\begin{itemize}
  \item a class $\Ob(\C)$ of \emph{objects};
  \item for every two objects $U,V\in\Ob(\C)$, a set $\Mor(U,V)$ of \emph{morphisms} from $U$ to $V$;
  for $A \in \Mor(U,V)$, we also write $A\colon U \to V$;
  \item for every three objects $U,V,W \in \Ob(\C)$, a binary operation $\circ\colon \Mor(V,W) \times \Mor(U,V) \to \Mor(U,W)$ called 
  \emph{composition of morphisms}, which is associative, that is,
	if $A\colon V \to W$, $B\colon U \to V$, $C\colon T \to U$, then $A \circ (B \circ C) = (A \circ B) \circ C$;
\end{itemize}
A semicategory $\C$ is called \emph{preadditive} if every set $\Mor(U,V)$ is an abelian group such that composition of morphisms is bilinear, that is,
	\begin{align*}
		A \circ (B + C) = (A \circ B) + (A \circ C) && \text{ and } && (A + B) \circ C = (A \circ C) + (B \circ C),
	\end{align*}
	where $+$ is the group operation.
\end{definition}

A semicategory can be thought of as a collection of objects, linked by arrows (the morphisms) that can be composed associatively.
Preadditive semicategories have the additional property that arrows with the same start and end can be added, yielding an abelian group structure that is compatible with the composition of morphisms.
Formally, a (preadditive) semicategory is just a (preadditive) category without identity morphisms.
For further information, see for example~\cite[Sec.~2]{garraway2005sheaves} or ~\cite[App.~B]{tilson}.
We also note that the words \emph{object} and \emph{morphism} do not imply anything about the nature of these things.
Intuitively, however, one can think of objects as sets and of morphisms as maps between those sets.

We list some classical examples of preadditive semicategories.

\begin{example}
In the following, $R$ denotes a ring (not necessarily with 1).
\begin{enumerate}
  \item The ring $R$ can be considered as a preadditive semicategory with only one object, and thus, only a single set of morphisms consisting of the underlying abelian group of $R$.
  Composition of morphisms is given by the ring multiplication.
  \item The set $\textbf{Mat}(R)$ of matrices with entries in $R$ can be considered as a preadditive category by taking as objects the 
  sets $R^n$ for all positive natural numbers $n$ and letting $\Mor(R^n,R^m) = R^{m \times n}$. 
  Composition is given by matrix multiplication.
  \item The category $R$-\textbf{Mod} of left modules over $R$ is a preadditive semicategory.
  Here, objects are left $R$-modules and morphisms are module homomorphisms between left $R$\nobreakdash-modules.  
  As a special case, we see that $K$-\textbf{Vect}, the category of vector spaces over a field $K$, is a preadditive semicategory.
  Note that the objects in these categories form proper classes and not sets.
  \item More generally, every preadditive category is a preadditive semicategory, thus so are, in particular, abelian categories.
\end{enumerate}
\end{example}

Using preadditive semicategories, we can summarise the discussion of this section in the following theorem.
In Section~\ref{sec:logical-framework}, we provide with Theorem~\ref{thm:idealisation} a theoretical justification for this conclusion.
In the following, we identify each identity of morphisms $P = Q$ with the noncommutative polynomial $p - q$ using the translation described above.
Furthermore, for noncommutative polynomials $f_1,\dots,f_r$, we denote by $(f_1,\dots,f_r)$ the (two-sided) ideal generated by $f_1,\dots,f_r$, consisting of all two-sided linear combinations of the $f_i$'s with polynomials as coefficients.

\begin{theorem}\label{thm:universal-statements-informal}
An identity $S = T$ of morphisms in a preadditive semicategory follows from other identities $A_1 =  B_1,\dots, A_n = B_n$ if and only if the ideal membership of noncommutative polynomials $s - t \in (a_1 - b_1,\dots,a_n - b_n)$ holds in the free algebra $\ZZ\<X>$.
\end{theorem}

Thus, based on Theorem~\ref{thm:universal-statements-informal} and the representation~\eqref{eq:proof-uniqueness-polynomials}, we can conclude this section with the following result concerning the uniqueness of the Moore-Penrose inverse.
To this end, we recall that a semicategory is \emph{involutive} if it is 
equipped with an involution $\ast$ that sends every morphism $A: U \to V$ to $A^\ast : V \to U$ and that satisfies~\eqref{eq:involution}.

\begin{theorem}
Let $A$ be a morphism in an involutive preadditive semicategory.
If there exist morphisms $B$ and $C$ satisfying~\eqref{eq:moore-penrose}, then $B = C$.
\end{theorem}

Thus far, our software package only supports computations in the free algebra $\QQ\<X>$.
To ensure that the computations are also valid over $\ZZ\<X>$, as required by Theorem~\ref{thm:universal-statements-informal}, one has to check whether all coefficients that appear in the computed cofactor representation are in fact integers.
The following routine \texttt{certify} builds a user-friendly wrapper around the ideal membership verification that also includes these checks.
It raises a warning if non-integer coefficients appear in the computed cofactor representation.

\begin{sageverbatim}
sage: F.<a, b, c, a_adj, b_adj, c_adj> = FreeAlgebra(QQ)
sage: Pinv_b = pinv(a, b, a_adj, b_adj)
sage: Pinv_c = pinv(a, c, a_adj, c_adj)
sage: assumptions = add_adj(Pinv_b + Pinv_c)
sage: proof = certify(assumptions, b-c)
Computing a (partial) Gröbner basis and reducing the claim...

Done! Ideal membership of all claims could be verified!
\end{sageverbatim}

\begin{remark}
We note that the computed $\texttt{proof}$ is the same as that computed by the $\texttt{ideal{\char`\_}membership}$ routine before.
\end{remark}

In many situations, all involved identities are of the form $P = Q$, where $P$ and $Q$ are compositions of basic operators or zero, as, for example, in case of the Penrose identities~\eqref{eq:moore-penrose}.
In such cases, all involved polynomials are binomials of the form $p - q$, where $p$ and $q$ are monomials in $\<X>$ or zero.
For these scenarios, \texttt{certify} is guaranteed to compute a cofactor representation with integer coefficients, provided that one exists.
However, for arbitrary polynomials, it could happen that \texttt{certify} only discovers a cofactor representation with rational coefficients, even if an alternative representation with integer coefficients exists.
We note, however, that in all the examples we have considered thus far, this situation has never occurred.

\section{Treating existential statements}
\label{sec:moore-penrose-existence}

Theorem~\ref{thm:universal-statements-informal} provides a method to verify whether an operator identity follows from other identities by checking ideal membership of noncommutative polynomials.
Although this technique is useful for proving various non-trivial statements, it still has its limitations.
Specifically, it does not cover existential statements that arise, for example, when solving operator equations.
This type of statement requires a slightly extended approach and cannot be proven solely by checking ideal membership.
In this section, we discuss how to treat existential statements.
As an illustrative example, we consider the existence of the Moore-Penrose inverse for complex matrices.
More precisely, we show that every complex matrix has a Moore-Penrose inverse, using polynomial computations.

In more general settings (e.g., bounded linear operators on Hilbert spaces,  $C^*$-algebras), not every element has a Moore-Penrose inverse.
Therefore, a crucial step in proving the desired statement is to characterise the fact that we are considering (complex) matrices.
In particular, since the polynomial framework can only deal with identities of operators, we have to express this fact in terms of identities.
One possibility to do this is via the singular value decomposition, which implies that, for every complex matrix $A$, there exist matrices $P,Q$ with
\begin{align}\label{eq:assumption-mp-existence}
          PA^*A = A&&\text{and}&&AA^*Q = A.
\end{align}
For example, if $A = U \Sigma V^*$ is a singular value decomposition of $A$, then $P = Q = U \Sigma^+ V^*$ is a possible choice, where $\Sigma^+$ is obtained from $\Sigma$ by replacing the non-zero diagonal entries by their reciprocals, and thus satisfies $\Sigma\Sigma^+ \Sigma^* = \Sigma^* \Sigma^+\Sigma = \Sigma$. 
Using this property of matrices, we can formalise the statement we consider in this section as the first-order formula
\begin{align*}
  \forall A,P,Q~\exists B \;:\; (PA^*A = A \;\wedge\; AA^*Q = A) \implies \eqref{eq:moore-penrose}.
\end{align*}

In the polynomial framework, the only possibility to prove such an existential statement is to derive an explicit expression for the existentially quantified objects.
Once such an explicit expression is obtained, the statement can be reformulated as a basic statement concerning identities, to which Theorem~\ref{thm:universal-statements-informal} can be applied.

For our example, this means finding an expression for $B$ in terms of $A,P,Q$ and their adjoints, modulo the assumptions~\eqref{eq:assumption-mp-existence}.
Algebraically, this corresponds to finding a polynomial $b = b(a,p,q,a^*,p^*,q^*)$ such that the elements~\eqref{eq:moore-penrose-polynomials}, representing the Penrose identities~\eqref{eq:moore-penrose}, lie in the ideal generated by 
\begin{align}\label{eq:assumption-mp-existence-polynomial} 
  &pa^* a - a,&& a a^* q - a,&& a^* a p^* - a^*,&& q^* a a^* - a^*,
\end{align}
encoding the assumptions~\eqref{eq:assumption-mp-existence}.

Through the use of Gröbner basis techniques, it is possible to employ a number of heuristics for finding elements of certain form in noncommutative polynomial ideals~\cite{heuristics}.
One such approach involves introducing a dummy variable $x$ for the desired expression $b$.
With this dummy variable, we consider the ideal $I$ generated by the assumptions (in our example given by~\eqref{eq:assumption-mp-existence-polynomial}) and by the identities that $b$ shall satisfy, but with $b$ replaced by $x$ (in our example these are the Moore-Penrose identities~\eqref{eq:moore-penrose-polynomials} for $x$).
Every polynomial of the form $x - b'$ in $I$ corresponds to a candidate expression $b'$ for $b$, and by applying the elimination property of Gröbner bases~\cite{borges1998groebner}, we can systematically search for such candidate expressions.
Our software package offers a user-friendly interface that simplifies the process of searching for expressions of this nature.

\begin{sageverbatim}
sage: F.<a,p,q,a_adj,p_adj,q_adj,x,x_adj> = FreeAlgebra(QQ)
sage: assumptions =  add_adj([a - p*a_adj*a, a - a*a_adj*q])
sage: Pinv_x = add_adj(pinv(a, x, a_adj, x_adj))
sage: I = NCIdeal(Pinv_x + assumptions)
sage: I.find_equivalent_expression(x)
\end{sageverbatim}
{{%
\abovedisplayskip=0pt plus 3pt 
\abovedisplayshortskip=0pt plus 3pt 
\begin{align*}
\text{\texttt{[- x + a{\char`\_}adj*q*x, - x + a{\char`\_}adj*p*x,}}\\
\text{\texttt{- x + a{\char`\_}adj*q*p{\char`\_}adj, -{ }x{ }+ a{\char`\_}adj*x{\char`\_}adj*x]}}
\end{align*}}}
Three out of the four candidate expressions for $b$ found by the heuristic still contain the dummy variable $x$ or its adjoint, and are thus useless.
However, the third polynomial $x - a^* q p^*$ shows that $b = a^* q p^*$ is a desired representation.
We use our software to show that $b$ satisfies the Moore-Penrose equations under the assumptions~\eqref{eq:assumption-mp-existence-polynomial}.

\begin{sageverbatim}
sage: MP_candidate = a_adj * q * p_adj
sage: MP_candidate_adj = p * q_adj * a
sage: claims = pinv(a, MP_candidate, a_adj, MP_candidate_adj)
sage: proof = certify(assumptions, claims)
Computing a (partial) Gröbner basis and reducing the claims...

Done! Ideal membership of all claims could be verified!
\end{sageverbatim}

We note that, here, \texttt{claims} is a list consisting of four polynomials, one for each of the four Penrose identities. 
In such cases, \texttt{certify} verifies the ideal membership of each element in the list and returns a list, here assigned to \texttt{proof}, providing a cofactor representation for each polynomial in \texttt{claims}.

Thus, we can conclude that every complex matrix $A$ has a Moore-Penrose inverse $A^\dagger$, given by $A^\dagger = A^* Q P^*$ with $P,Q$ as in~\eqref{eq:assumption-mp-existence}.
More generally, by Theorem~\ref{thm:universal-statements-informal}, we have proven the following statement.

\begin{theorem}\label{thm:moore-penrose-existence}
Let $A$ be a morphism in an involutive preadditive semicategory.
If there exist morphisms $P$ and $Q$ satisfying~\eqref{eq:assumption-mp-existence}, then $A$ has a Moore-Penrose inverse $A^\dagger$, given by $A^\dagger = A^* Q P^*$.
\end{theorem}

\begin{remark}
Typically, under the assumptions~\eqref{eq:assumption-mp-existence} the Moore-Penrose inverse is expressed by the formula $A^\dagger = Q^* A P^*$, cf.~\cite[Lem.~3]{PR81}.
We note that this expression is equivalent to ours, and can be found using our software by changing the monomial order underlying the polynomial computation.
\begin{sageverbatim}
sage: I.find_equivalent_expression(x,
....: order=[[q,q_adj,a,a_adj,p,p_adj],[x,x_adj]])[0]
\end{sageverbatim}
{{%
\abovedisplayskip=0pt plus 3pt 
\abovedisplayshortskip=0pt plus 3pt 
\begin{displaymath}
\text{\texttt{- q{\char`\_}adj*a*p{\char`\_}adj + x}}
\end{displaymath}}}
This gets to show that the output of the polynomial heuristics depends strongly on several parameters, and in particular, on the used monomial order.
\end{remark}

We could prove Theorem~\ref{thm:moore-penrose-existence} by explicitly constructing an expression for the existentially quantified operator.
This now raises the question whether this is always possible or whether we just got lucky in this example.
\emph{Herbrand's theorem}~\cite{Her30,buss1994herbrand}, a fundamental result in mathematical logic, provides an answer to this question.
It states that such an explicit representation always exists and can be constructed as a polynomial expression in terms of the basic operators appearing in the statement, provided that the operator statement is true in every preadditive semicategory.
We refer to Theorem~\ref{thm:herbrand} for the precise statement.
Thus, by enumerating all such polynomial expressions, we are guaranteed to find a correct instantiation if the considered statement is correct.

Of course, naively enumerating all possible polynomial expressions quickly becomes infeasible.
Therefore, it is important to have good heuristics that allow to systematically search for suitable candidate expressions.
Our software package implements, apart from the heuristic described above, several such techniques for finding polynomials of special form in noncommutative ideals.
We refer to Appendix~\ref{appendix} for further information and the corresponding commands.
Most importantly, it provides methods for finding factorisations of given operators.
This allows to effectively model many properties of operators, including conditions on ranges and kernels, as well as injectivity and surjectivity, or more generally, cancellability properties.
In the following section, we discuss how properties like these can be treated within the framework.

\section{Treating common properties}
\label{sec:common-properties}

\subsection{Real matrices}

A property that appears regularly in matrix statements, especially in combination with the Hermitian adjoint $A^*$, is that of having matrices over the reals.
It can be encoded by decomposing the Hermitian adjoint $A^*$ into an entry-wise complex conjugation, denoted by $A^C$, followed by a transposition, that is, $A^* = (A^C)^T$.
With this, a matrix $A$ being real can be expressed algebraically by the identity $A = A^C$, exploiting the fact that the conjugate of a real number is the number itself.

To model the complex conjugation and the transposition on the polynomial level, we proceed analogous to modelling the involution (see Remark~\ref{remark:involution}).
We introduce additional variables $a^C$  and $a^T$ for the complex conjugate and the transpose, respectively, of each basic operator $A$.
Additionally, for every assumption $P = Q$, we have to translate, next to the corresponding adjoint identity $P^* = Q^*$, now also the transposed identity $P^T = Q^T$ as well as the conjugated identity $P^C = Q^C$ into polynomials.
These additional identities first have to be simplified using the following rules that relate the different function symbols to each other.
\begin{align*}
  (P+Q)^\alpha &= P^\alpha + Q^\alpha, & (P^\alpha)^\beta &= \begin{cases} 
                                                              P & \text{ if } \alpha = \beta \\ 
                                                              P^\gamma & \text{ if } \alpha \neq \beta
                                                            \end{cases},\\
  (PQ)^C &= P^C Q^C, & (PQ)^\delta &= Q^\delta P^\delta,
\end{align*}
with $\alpha,\beta,\gamma,\delta \in \{\ast, C, T\}$ such that $\gamma \neq \alpha, \beta$ and $\delta \neq C$.

As an illustrative example, we consider the statement that the Moore-Penrose inverse $B$ of a real matrix $A$ is real as well.
With the help of our software package, it can be proven as follows.

\begin{sageverbatim}
sage: F.<a, a_tr, a_c, a_adj, b, b_tr, b_c, b_adj> = FreeAlgebra(QQ)

# the basic assumptions
sage: Pinv_b = add_adj(pinv(a, b, a_adj, b_adj))
# the transposed and conjugated assumptions
sage: Pinv_b_tr = [a_tr*b_tr*a_tr - a_tr, b_tr*a_tr*b_tr - b_tr,
....: a_tr*b_tr - b_c*a_c, b_tr*a_tr - a_c*b_c]
sage: Pinv_b_c = [a_c*b_c*a_c - a_c, b_c*a_c*b_c - b_c]
# assumption that a is real
sage: a_real = [a - a_c, a_tr - a_adj]

sage: assumptions = Pinv_b + Pinv_b_tr + Pinv_b_c + a_real
sage: proof = certify(assumptions, b - b_c)
Computing a (partial) Gröbner basis and reducing the claims...

Done! Ideal membership of all claims could be verified!
\end{sageverbatim}

\subsection{Identity operators}

Next, we discuss how to handle identity matrices or operators.
While zero operators have a natural translation into the zero polynomial, identity operators cannot be directly mapped to the multiplicative identity $1$ in the free algebra, as this would constitute a many-to-one mapping and a loss of information.
We note that this is not an issue when mapping all zero operators to the zero polynomial, as the zero polynomial does not affect any polynomial computations.

Instead, identity operators have to be treated like any other basic operator,
which means introducing a new indeterminate $i_u$ for every identity operator $I_U$ and explicitly adding the identities satisfied by $I_U$ to the assumptions.
In particular, these are the idempotency of $I_U$, the fact that $I_U$ is self-adjoint, and the identities $A I_U = A$ and $I_U B = B$ for all basic operators $A,B$ for which these expressions are well-defined. 

We illustrate the handling of identity operators in the next section.

\subsection{Injectivity, surjectivity, and full matrix ranks}

Injectivity and surjectivity of operators appear regularly as properties in statements.
They can be encoded by exploiting the following classical fact.
\begin{lemma}\label{lemma:injective-surjective}
Let $U,V$ be non-empty sets.
A function $A: U \to V$ is
\begin{enumerate}
  \item injective if and only if $A$ has a left inverse $B : V \to U$;
  \item surjective if and only if $A$ has a right inverse $C : V \to U$;
\end{enumerate}
\end{lemma}

Thus, an assumption of injectivity of an operator $A$ can be encoded via the identity $BA = I_U$, where $B$ is a new operator that does not satisfy any additional hypotheses and $I_U$ is the identity on $U$.
Analogously, surjectivity of $A$ corresponds to the identity $AC = I_V$.
For proving injectivity or surjectivity of an operator in our setting, we have to show the existence of a left or right inverse by finding an explicit expression for such an operator.

As a special case of the discussion above, we also obtain a way to encode the property of a matrix having full row or column rank.
This follows from the fact that a matrix $A$ has full row rank if and only if the associated linear function is surjective, which, by Lemma~\ref{lemma:injective-surjective}, is the case if and only if $A$ has a right inverse.
Dually, $A$ has full column rank if and only if $A$ has a left inverse.

To illustrate the handling of full rank assumptions as well as of identity matrices, we consider the statement:
If $A = BC$ is a full rank decomposition of a matrix $A$, i.e., $B$ has full column rank and $C$ has full row rank, then $A^\dagger = C^\dagger B^\dagger$.
Using the software, it can be proven as follows.

\begin{sageverbatim}
sage: F.<a, b, c, i, u, v, x, y, z, a_adj, b_adj, c_adj, i_adj, 
....: u_adj, v_adj, x_adj, y_adj, z_adj> = FreeAlgebra(QQ)

sage: Pinv_a = pinv(a, x, a_adj, x_adj)
sage: Pinv_b = pinv(b, y, b_adj, y_adj)
sage: Pinv_c = pinv(c, z, c_adj, z_adj)
# full ranks encoded via one-sided inverses
sage: rank_decomp = [a - b*c, u*b - i, c*v - i]
# encode identity i
sage: id = [i*i - i, i - i_adj, b*i - b, i*y - y, i*c - c, 
....: z*i - z, i*u - u, v*i - v]

sage: assumptions = add_adj(Pinv_a + Pinv_b + Pinv_c +
....: rank_decomp + id)
sage: claim = x - z*y
sage: proof = certify(assumptions, claim)
Computing a (partial) Groebner basis and reducing the claims...

Starting iteration 5...
Done! Ideal membership of all claims could be verified!
\end{sageverbatim}

\begin{remark}
We note that the \texttt{certify} routine (more precisely, the Gr\"obner basis computation underlying this command) is an iterative procedure.
By default, the package informs about the computational progress of this procedure by printing an update message \texttt{Starting iteration n...} every fifth iteration, see also Section~\ref{sec:certify}.
\end{remark}

\subsection{Range inclusions}

Another common class of properties are conditions on ranges and kernels, like the inclusion of ranges $R(A) \subseteq R(B)$ of operators $A, B$.
In case of linear operators over a field, such a range inclusion can be translated into the existence of a factorisation $A = BX$ for some operator $X$.
We note that, in Hilbert and Banach spaces, this is the well-known factorisation property in Douglas' lemma~\cite{douglas}.

Thus, also facts like $R(A^\dag) = R(A^*)$ can be treated within the framework by finding explicit factorisations of $A^\dag$ and $A^*$ in terms of the other. 
Using our software, such factorisations can be found easily.
We refer to Section~\ref{appendix:heuristics} for the available heuristics to do this, and to~\cite{heuristics} for a more thorough explanation of these techniques.

\begin{sageverbatim}
sage: F.<a, a_adj, a_dag, a_dag_adj> = FreeAlgebra(QQ)
sage: Pinv_a = add_adj(pinv(a, a_dag, a_adj, a_dag_adj))
sage: I = NCIdeal(Pinv_a)
# R(A^\dag) \subseteq R(A^*)
sage: I.find_equivalent_expression(a_dag, prefix=a_adj,
....: heuristic='naive')
\end{sageverbatim}
{{%
\abovedisplayskip=0pt plus 3pt 
\abovedisplayshortskip=0pt plus 3pt 
\begin{align*}
\text{\texttt{[- a{\char`\_}dag + a{\char`\_}adj*a{\char`\_}dag{\char`\_}adj*a{\char`\_}dag]}}
\end{align*}}}
\begin{sageverbatim}
# R(A^*) \subseteq R(A^\dag)
sage: I.find_equivalent_expression(a_adj, prefix=a_dag,
....: heuristic='naive')
\end{sageverbatim}
{{%
\abovedisplayskip=0pt plus 3pt 
\abovedisplayshortskip=0pt plus 3pt 
\belowdisplayskip=0pt minus 3pt 
\belowdisplayshortskip=0pt minus 3pt 
\begin{align*}
\text{\texttt{[- a{\char`\_}adj + a{\char`\_}adj*a*a{\char`\_}dag]}}
\end{align*}}}

\section{Logical framework}
\label{sec:logical-framework}

In the following, we describe the theory developed in~\cite{hofstadler2022universal} from a practical point of view, focusing on $\forall \exists$-statements.
For the reduction from arbitrary first-order formulas to this case, as well as for all proofs and additional resources, we refer to the corresponding sections in~\cite{hofstadler2022universal}.

To model statements about linear operators, or more generally about morphisms in preadditive semicategories, we consider a subset of many-sorted first-order logic.
Many-sorted first-order logic extends classical first-order logic by assigning a sort to each term.
These sorts allow to represent objects from different universes and restrict which expressions can be formed.
In our context, they are used to represent domains and codomains of operators.

To formally introduce operator statements, we fix an enumerable set of 
\emph{object symbols} $\ObS = \{v_1,v_2,\dots \}$.
We call a pair $(u,v) \in \ObS \times \ObS$ a \emph{sort}.
We also fix an enumerable set of variables $\{x_1,x_2,\dots\}$ as well as, for each sort $(u,v)$, a \emph{zero constant} $0_{u,v}$.
Furthermore, we fix a \emph{sort function} $\sigma$ mapping each variable $x$ to a sort $\sigma(x) \in \ObS \times \ObS$ and each zero constant $0_{u,v}$ to $\sigma(0_{u,v}) = (u,v)$.
Intuitively, variables correspond to basic operators and the zero constants model distinguished zero operators.
The images of these symbols under the sort function $\sigma$ represent their domains and codomains.

Using these basic symbols, we can construct terms, and building upon that, operator statements.
Note that the following definition also extends the sort function from variables and constants to terms.

\begin{definition}
A \emph{term} is any expression that can be built up inductively using the following rules:
\begin{enumerate}
  \item each variable $x$ is a term of sort $\sigma(x)$;
  \item each zero constant $0_{u,v}$ is a term of sort $(u,v)$;
  \item if $s,t$ are terms of sort $\sigma(s) = \sigma(t)$, then $s+t$ is a term of sort $\sigma(s+t) \coloneqq \sigma(s)$;
  \item if $s,t$ are terms of sort $\sigma(s) = (v,w)$, $\sigma(t) = (u,v)$, then $st$ is a term of sort $\sigma(st) \coloneqq (u,w)$;
\end{enumerate}
\end{definition}

Terms are simply all noncommutative polynomial expressions that can be formed from the variables and the zero constants under the restrictions imposed by the sort function.
They correspond to all operators that can be formed from the basic operators with the arithmetic operations of addition and composition.

\begin{definition}\label{def:operator-statement}
An \emph{operator statement} is a first-order formula that can be built up inductively using the following rules:
\begin{enumerate}
  \item if $s,t$ are terms of sort $\sigma(s) = \sigma(t)$, then $s = t$ is an operator statement;
   \item if $\varphi$ is an operator statement, then so is $\lnot \varphi$;
  \item if $\varphi, \psi$ are operator statements, then so is $\varphi \ast \psi$ for $\ast \in \{\vee, \wedge, \rightarrow\}$;
  \item if $\varphi$ is an operator statement, then so is $P x: \varphi$ for any variable $x$ and $P \in \{\exists,\forall\}$;
\end{enumerate}
\end{definition}

\begin{remark}
We consider $\wedge$ and $\vee$ as associative and commutative operations, i.e., $\varphi \wedge (\psi \wedge \rho) = (\varphi \wedge \psi) \wedge \rho$ and $\varphi \wedge \psi = \psi \wedge \varphi$, and analogously for $\vee$.
Furthermore, we abbreviate $\lnot (s = t)$ by $s \neq t$ in the following.
\end{remark}

We recall some standard definitions and notation.
In the last point of Definition~\ref{def:operator-statement}, $P$ is called the \emph{quantifier} of $x$ and $\varphi$ is the \emph{scope} of $Px$.
If all variables occurring in an operator statement $\varphi$ are in the scope of a quantifier, then $\varphi$ is \emph{closed}.
We abbreviate a block of consecutive equally quantified variables $Px_1Px_2\dots Px_k$, with $P \in \{\exists,\forall\}$, by $Px_1\dots x_k$, or simply by $P\mathbf{x}$.
Furthermore, to indicate the scope of a quantifier, we also write $P \mathbf{x} : \varphi(\mathbf{x})$.

An operator statement without any quantifiers is called \emph{quantifier-free}.
Moreover, any operator statement of the form $\forall \mathbf{x} : \varphi(\mathbf{x})$ (resp.~$\exists \mathbf{x} : \varphi(\mathbf{x})$) with $\varphi$ quantifier-free is called \emph{universal} (resp.~\emph{existential}), and any operator statement of the form $\forall \mathbf{x} \exists \mathbf{y} : \varphi(\mathbf{x},\mathbf{y})$ with $\varphi$ quantifier-free is a $\forall\exists$-operator statement.

An \emph{interpretation} $\mathcal{I}$ allows to interpret an operator statement $\varphi$ as a statement about morphisms in a preadditive semicategory $\mathcal{C}$.
It assigns to each object symbol $u \in \ObS$ an object $\mathcal{I}(u) \in \Ob(\mathcal{C})$ and to each variable $x$ of sort $\sigma(x) = (u,v)$ a morphism $\mathcal{I}(x) \colon \mathcal{I}(u) \to \mathcal{I}(v)$.
Each zero constant $0_{u,v}$ is mapped to the zero morphism in the abelian group $\Mor(\mathcal{I}(u),\mathcal{I}(v))$.
This ensures that the terms in $\varphi$ are translated into well-formed morphisms in $\mathcal{C}$.
Then $\varphi$ can be evaluated to a truth value by interpreting the boolean connectives and the quantifiers like in classical first-order logic.

\begin{definition}
An operator statement $\varphi$ is \emph{universally true} if $\varphi$ evaluates to true under all possible interpretations in every preadditive semicategory $\mathcal{C}$.
\end{definition}

Note that an interpretation of $\varphi$ depends implicitly on the sort function $\sigma$, and thus, so does the semantic evaluation of $\varphi$.
An operator statement may be universally true w.r.t.\ one sort function but not w.r.t.\ another sort function.
For instance, statements that hold for square matrices may not hold for rectangular matrices.
Therefore, we should only refer to universal truth w.r.t.\ a specific sort function.
For the sake of brevity, we assume a fixed sort function $\sigma$ and disregard this dependency in the following.

\begin{remark}
For a formal definition of interpretation and universal truth of operator statements, we refer to~\cite[Sec.~2.1.2]{hofstadler2022universal}.
\end{remark}

In the remainder of this section, we characterise universal truth of operator statements by ideal membership of noncommutative polynomials.
To this end, we recall that every quantifier-free operator statement $\varphi$ can be transformed into a logically equivalent formula of the form
\begin{align}
\label{eq:cnf}
	\bigwedge_{i=1}^{m} \left( \bigvee_{j=1}^{n_i} a_{i,j} \neq b_{i,j} \vee \bigvee_{k=1}^{n'_i} s_{i,k} = t_{i,k} \right).
\end{align}

In the above formula, either of the two disjunctions can also be empty, i.e., it is possible that either $n_i = 0$ or $n_i' = 0$, but not both.

We recall that a formula of the form~\eqref{eq:cnf} is in \emph{conjunctive normal form (CNF)}~\cite{Pre09}.
It is a conjunction of \emph{clauses}, where a clause is a disjunction of equalities and disequalities.
A formula can have several CNFs.
One way to obtain a CNF of a quantifier-free operator statement $\varphi$ is to apply to $\varphi$ exhaustively each of the following sets of rewrite rules, in the given order:
\begin{enumerate}
	\item Eliminate implications: $\qquad\psi_1 \rightarrow \psi_2 \;\rightsquigarrow\; \lnot \psi_1 \vee \psi_2$
	\item Move $\lnot$ inwards (i.e., compute a negation normal form):
	\begin{align*}
\lnot \lnot \psi  \;\rightsquigarrow\; \psi  \qquad
	\lnot (\psi_1 \wedge \psi_2) \;\rightsquigarrow\; \lnot \psi_1 \vee \lnot \psi_2 \qquad
	\lnot (\psi_1 \vee \psi_2) \;\rightsquigarrow\; \lnot \psi_1 \wedge \lnot \psi_2
	\end{align*}
	\item Distribute $\vee$ over $\wedge$: $\qquad\psi \vee (\psi_1 \wedge \psi_2) \;\rightsquigarrow\; (\psi \vee \psi_1) \wedge (\psi \vee \psi_2)$
\end{enumerate}

We note that the above rules apply modulo associativity and commutativity of $\wedge, \vee$.
This process yields a unique normal form, which we denote by $\CNF(\varphi)$.
Also note that this transformation preserves the semantics of $\varphi$, that is, $\CNF(\varphi)$ is logically equivalent to $\varphi$.

Based on the conjunctive normal form, we define a translation of operator statements into ideal theoretic statements. This process is called \emph{idealisation}.
We first discuss the special case of clauses.
To this end, we associate to each equality $s = t$ or disequality $s \neq t$ of terms the noncommutative polynomial $s - t$ using the same translation as for identities of operators described in Section~\ref{sec:moore-penrose-uniqueness}.

\begin{definition}
Let $C = \bigvee_{j=1}^{n} a_{j} \neq b_{j} \vee \bigvee_{k=1}^{n'} s_k = t_k$ be a clause.
The \emph{idealisation} $\Id(C)$ of $C$ is the following predicate considered as a statement in the free algebra $\ZZ\<X>$:
\begin{align*}
	\Id(C) \; :\equiv \; s_{k} - t_{k} \in \left(a_{1} - b_{1},\dots, a_n - b_n\right) \text{ for some } 1 \leq k \leq n'.
\end{align*}
\end{definition}

To motivate this definition, write $C$ in the equivalent form $\bigwedge_j a_{j} = b_{j} \rightarrow \bigvee_k s_{k} = t_{k}$.
This shows that $C$ is true if and only if at least one of the identities $s_{k} = t_{k}$ can be derived from all the $a_{j} = b_{j}$.
Precisely this fact is described by $\Id(C)$.

The process of idealisation extends to universal operator statements as follows.

\begin{definition}
Let $\varphi = \forall \mathbf{x} : \psi(\mathbf{x})$ be a universal operator statement. 
The \emph{idealisation} $\Id(\varphi)$ of $\varphi$ is the predicate
\begin{align*}
	\Id(\varphi) :\equiv \bigwedge_{\substack{C \text{ clause}\\[0.1em] \text{of } \CNF(\psi)}} \Id(C).
\end{align*}
\end{definition}

The following theorem links the universal truth of universal operator statements to their idealisation.

\begin{theorem}[{\cite[Thm.~27]{hofstadler2022universal}}]\label{thm:idealisation}
A universal operator statement $\varphi$ is universally true if and only if the idealisation of $\varphi$ is true.
\end{theorem}

\begin{remark}
Theorem~\ref{thm:idealisation} is a formalisation of the informal  
Theorem~\ref{thm:universal-statements-informal}.
It is a generalisation of~\cite[Thm.~32]{RRH21}, which only considers a 
restricted form of universal operator statements and only provides a
sufficient condition for the universal truth of the operator statement.
\end{remark}

Theorem~\ref{thm:idealisation} reduces universal truth of universal operator statements to the verification of finitely many polynomial ideal memberships. 
Since the latter problem is semi-decidable, this immediately yields a semi-decision procedure for universal truth of this kind of statements.

In the following, we describe how to treat operator statements involving existential quantifiers.
Although the subsequent results can be phrased for arbitrary first-order formulas, we focus on the more practical and important case of closed $\forall\exists$-operator statements.
It is worth noting that any operator statement can be transformed into a logically equivalent formula of this form.
For more information on this conversion, we refer to~\cite[Sec.~2.2 and~2.3]{hofstadler2022universal}.

The following result is an adaptation of one of the most fundamental theorems of mathematical logic, \emph{Herbrand's theorem}~\cite{Her30}, to our setting.
It essentially allows to eliminate existential quantifiers, reducing the treatment of $\forall\exists$-operator statements to universal ones.
To state the theorem, we recall the concept of Herbrand expansion.
The \emph{Herbrand expansion} $H(\varphi)$ of a closed $\forall\exists$-operator statement $\varphi = \forall \mathbf{x} \exists y_1,\dots,y_k: \psi(\mathbf{x}, \mathbf{y})$ is the set of all instantiations of the existentially quantified variables of $\varphi$, that is,
\begin{align*}
  H(\varphi) \coloneqq \left\{\psi(\mathbf{x},t_1,\dots,t_k) \mid t_i \text{ terms only involving }\mathbf{x}\text{ s.t. }\sigma(t_i) = \sigma(y_i)\right\}.
\end{align*}

We note that $H(\varphi)$ is an infinite yet enumerable set of quantifier-free operator statements.

\begin{theorem}\label{thm:herbrand}
A closed $\forall\exists$-operator statement $\varphi = \forall \mathbf{x} \exists \mathbf{y}: \psi(\mathbf{x}, \mathbf{y})$ is universally true if and only if there exist finitely many $\varphi_1(\mathbf{x}),\dots,\varphi_n(\mathbf{x}) \in H(\varphi)$ such that the universal operator statement
$\forall \mathbf{x} : \bigvee_{i=1}^n \varphi_i(\mathbf{x})$
is universally true.
\end{theorem}

\begin{remark}
Theorem~\ref{thm:herbrand} is a special case of the more general~\cite[Cor.~15]{hofstadler2022universal}.
It can also be found in~\cite{buss1994herbrand} for classical unsorted first-order logic.
\end{remark}

The following steps give an overview on how Herbrand's theorem can be used algorithmically to reduce the treatment of a closed $\forall\exists$-operator statement $\varphi$ to a universal one.
They can be considered as an adaptation of Gilmore's algorithm~\cite{gilmore1960proof}.

\begin{enumerate}
	\item Let $\varphi_1,\varphi_2,\dots$ be an enumeration of $H(\varphi)$.
	\item Let $n = 1$.
	\item Form the formula $\psi_n = \forall \mathbf{x}: \bigvee_{i=1}^n \varphi_i(\mathbf{x})$.\label{step loop}
	\item If the idealisation $I(\psi_n)$ is true, then $\varphi$ is universally true.
	Otherwise, increase $n$ by $1$ and go to step~\ref{step loop}.\label{step check}
\end{enumerate}

Since first-order logic is only semi-decidable, we cannot expect to obtain an algorithm that terminates on any input.
The best we can hope for is a semi-decision procedure that terminates if and only if an operator statement $\varphi$ is universally true.
However, the steps above, as phrased now, still have a subtle flaw that stops them from even being a semi\nobreakdash-decision procedure. 

The conditional check in step~\ref{step check} requires to decide certain ideal memberships.
While \emph{verifying} ideal membership of noncommutative polynomials is always possible in finite time, \emph{disproving} it is generally not.
Consequently, verifying that the condition in step~\ref{step check} is false is generally not possible in finite time.
In cases where this is required, the procedure cannot terminate -- even if $\varphi$ is indeed universally true.

To overcome this flaw and to obtain a true semi-decision procedure, we have to interleave the computations done for different values of $n$.
Procedure~\ref{proc:procedure} shows one way how this can be done.
It essentially follows the steps described above, except that it only performs finitely many operations to check if $I(\psi_n)$ is true for each $n$.

\begin{algorithm}[H]
\renewcommand{\algorithmcfname}{Procedure}
\Input{A closed $\forall\exists$-operator statement $\varphi$}
\Output{\texttt{true} if and only if $\varphi$ is universally true; otherwise infinite loop}

$\varphi_1,\varphi_2,\ldots \leftarrow$ an enumeration of $H(\varphi)$\;
\For{$n \leftarrow 1,2,\dots$}{
  	$\psi_{n} \leftarrow \forall \mathbf{x} : \bigvee_{i=1}^{n} \varphi_i(\mathbf{x})$
    \label{line instance}\;
  \For{$k \leftarrow 1,\dots,n$}{
  \If{$I(\psi_{k}) = \texttt{true}$ can be verified with $n$ operations of an ideal membership semi\nobreakdash-decision procedure\label{line ideal checks}}
  { 
    \Return{\texttt{true}}\;
   }
   }
 }
\caption{Semi-decision procedure for verifying operator statements}
\label{proc:procedure}
\end{algorithm}

Line~\ref{line ideal checks} of Procedure~\ref{proc:procedure} contains the term \emph{operation of a procedure}.
Thereby we mean any (high- or low-level) set of instructions of the procedure that can be executed in \emph{finite time}.

\begin{theorem}[{\cite[Thm.~36]{hofstadler2022universal}}]\label{thm:semi-decision-procedure}
Let $\varphi$ be a closed $\forall\exists$-operator statement.
Procedure~\ref{proc:procedure} terminates and outputs \texttt{true} if and only if $\varphi$ is universally true.
\end{theorem}

\begin{remark}
Procedure~\ref{proc:procedure} is a special case of the more general semi-decision procedure~\cite[Proc.~2]{hofstadler2022universal} that allows to treat arbitrary operator statements.
\end{remark}



\section{Case Study}
\label{sec:case-study}
For our case study, we considered the first 25 facts in the section on the Moore-Penrose inverse in the Handbook of Linear Algebra~\cite[Sec.~I.5.7]{hogben2013handbook}.
Among these 25 statements, we found that five cannot be treated within the framework, as they contain properties that cannot be expressed in terms of identities of operators (e.g., properties of the matrix entries, norms, or statements that require induction).
Additionally, three statements can only be partially handled for the same reason.
The remaining 17 statements, along with those parts of the three statements mentioned before that can be treated within the framework, can all be translated into polynomial computations and proven fully automatically with the help of our software.
The corresponding polynomial computations take place in ideals generated by up to 70 polynomials in up to 18 indeterminates.
The proof of each statement takes less than one second and the computed cofactor representations, certifying the required ideal memberships, consist of up to 226 terms.

As part of our case study, we also examined Theorems~2.2 -- 2.4 in~\cite{DD10}, which provide several necessary and sufficient conditions for the reverse order law $(AB)^\dag = B^\dag A^\dag$ to hold, where $A,B$ are bounded linear operators on Hilbert spaces with closed ranges.
Our software can automatically prove all of these statements in less than five seconds altogether, yielding algebraic proofs that consist of up to 279 terms.
We note that, in contrast to the original proofs in~\cite{DD10}, which rely on matrix forms of bounded linear operators that are induced by some decompositions of Hilbert spaces, our proofs do not require any structure on the underlying spaces except basic linearity and a certain cancellability assumption.
This implies that our proofs generalise the results from bounded operators on Hilbert spaces to morphisms in arbitrary preadditive semicategories, meeting the cancellability requirement.


Finally, our case study contains fully automated proofs of Theorem~2.3 and~2.4 in the recent paper~\cite{cvetkovic2021algebraic}, which provide necessary and sufficient conditions for the triple reverse order law $(ABC)^\dag = C^\dag B^\dag A^\dag$ to hold, where $A,B,C$ are elements in a ring with involution.
These results, which provide several improvements of Hartwig's classical triple reverse order law~\cite{Har86}, were motivated and partly discovered by a predecessor of our software package~\cite{HRR19}. Our new software can automate all aspects of the proofs, relying heavily on the heuristics for finding polynomials of special form in ideals.
We note that, while an initial implementation of our package took several days to complete the computations required for proving these theorems, the version discussed here now performs the task in approximately 15 seconds.
The assumptions in the proof of Theorem~2.3 consist of up to 24 polynomials in 22 indeterminates and the computed cofactor representations certifying the ideal membership have up to 80 terms.
The software also allows for easy experimentation with relaxing the assumptions of a theorem. 
This led us, among other simplifications, to discover that a condition in the original theorem~\cite{Har86} requiring equality of certain ranges $R(A^*AP) = R(Q^*)$ can be replaced with the weaker condition of a range inclusion $R(A^*AP) \subseteq R(Q^*)$.

\subsubsection{Acknowledgements}

We thank the anonymous referees for their careful reading and valuable suggestions which helped to improve the presentation of this work.

\bibliographystyle{splncs04}
\bibliography{references}

\appendix
\section{The software package \texttt{operator\char`_gb}}
\label{appendix}

In this appendix, we give an introduction to the functionality provided by the \textsc{SageMath} package \texttt{operator\char`_gb} for Gröbner basis computations in the free algebra, with a particular focus on methods that facilitate proving statements about linear operators.
We assume that readers are already familiar with \textsc{SageMath} and, for reading Section~\ref{appendix:ideal-membership} and~\ref{appendix:heuristics}, with the theory of Gröbner bases in the free algebra.
For further information on these topics, we refer to~\cite{sagemath} and to~\cite{Xiu12,Mor16}, respectively.

At the time of writing, the package is still under development and not part of the official \textsc{SageMath} distribution.
The current version, however, can be downloaded from
\begin{center}
  \url{https://github.com/ClemensHofstadler/operator_gb}
\end{center}
and installed as described on the webpage.
The code can then be loaded into a \textsc{SageMath} session by the following command.
\begin{sageverbatim}
sage: from operator_gb import *
\end{sageverbatim}

For now, the package only offers functionality for computations over the coefficient domain $\mathbb{Q}$.
In the future, we will extend the functionality to other (finite) fields and subsequently also to coefficient rings such as $\mathbb{Z}$.
Furthermore, we also plan on integrating noncommutative signature-based Gröbner algorithms~\cite{hofstadler-verron} and, based on them, newly developed methods to compute cofactor representations of minimal length~\cite{hofstadler2023short}.

\subsection{Certifying operator statements}
\label{sec:certify}

The basic use-case of the package is to compute proofs of operator statements by certifying ideal membership of noncommutative polynomials.
To this end, the package provides the command \texttt{certify(assumptions, claim)}, which allows to certify whether a noncommutative polynomial \texttt{claim} lies in the ideal generated by a list of polynomials \texttt{assumptions}.

For example, to certify that $abc - d$ lies in the ideal generated by $ab - d$ and $c - 1$, proceed as follows.
\begin{sageverbatim}
sage: F.<a,b,c,d> = FreeAlgebra(QQ)
sage: assumptions = [a*b - d, c - 1]
sage: proof = certify(assumptions, a*b*c - d)
Computing a (partial) Groebner basis and reducing the claims...

Done! Ideal membership of all claims could be verified!
\end{sageverbatim}

\begin{remark}
Note that noncommutative polynomials are entered using the \texttt{FreeAlgebra} data structure provided by \textsc{SageMath}.
\end{remark}

The computed \texttt{proof} provides a cofactor representation of \texttt{claim} in terms of the elements in \texttt{assumptions}.
More precisely, it is a list of tuples $(a_{i}, j_i, b_i)$ with terms $a_i, b_i$ in the free algebra and integers $j_i$ such that 
\[
  \mathtt{claim} \;= \sum_{i=1}^{|\mathtt{proof}|} a_i \cdot \mathtt{assumptions}[j_i] \cdot b_i.
\]
The package provides a \texttt{pretty\char`\_print\char`\_proof} command to visualise the proof in form of a string.
It also allows to expand a cofactor representation using the command \texttt{expand\_cofactors}.

\begin{sageverbatim}
sage: proof
\end{sageverbatim}
{{%
\abovedisplayskip=0pt plus 3pt 
\abovedisplayshortskip=0pt plus 3pt 
\begin{displaymath}
\text{\texttt{[(1,0,c), (d,1,1)]}}
\end{displaymath}}}
\begin{sageverbatim}
sage: pretty_print_proof(proof, assumptions)
\end{sageverbatim}
{{%
\abovedisplayskip=0pt plus 3pt 
\abovedisplayshortskip=0pt plus 3pt 
\begin{displaymath}
\text{\texttt{-d + a*b*c = (-d + a*b)*c + d*(-1 + c)}}
\end{displaymath}}}
\begin{sageverbatim}
sage: expand_cofactors(proof, assumptions)
\end{sageverbatim}
{{%
\abovedisplayskip=0pt plus 3pt 
\abovedisplayshortskip=0pt plus 3pt 
\belowdisplayskip=0pt 
\belowdisplayshortskip=0pt
\begin{displaymath}
\text{\texttt{-d + a*b*c}}
\end{displaymath}}}
\begin{remark}
The \texttt{certify} command also checks if the computed cofactor representation is valid over $\mathbb{Z}$ as well, i.e., if all coefficients that appear are integers.
If this is not the case, it produces a warning, but still continues the computation and returns the result.
\end{remark}

It is also possible to give \texttt{certify} a list of polynomials as \texttt{claim}.
In this case, a cofactor representation of each element in \texttt{claim} is computed.
\begin{sageverbatim}
sage: claims = [a*b*c - d, a*b - c*d]
sage: proof = certify(assumptions, claims)
Computing a (partial) Groebner basis and reducing the claims...

Done! Ideal membership of all claims could be verified!
sage: pretty_print_proof(proof[0], assumptions)
\end{sageverbatim}
{{%
\abovedisplayskip=0pt plus 3pt 
\abovedisplayshortskip=0pt plus 3pt 
\belowdisplayskip=0pt plus 3pt 
\belowdisplayshortskip=0pt plus 3pt 
\begin{displaymath}
\text{\texttt{-d + a*b*c = (-d + a*b)*c + d*(-1 + c)}}
\end{displaymath}}}
\begin{sageverbatim}
sage: pretty_print_proof(proof[1], assumptions)
\end{sageverbatim}
{{%
\abovedisplayskip=0pt plus 3pt 
\abovedisplayshortskip=0pt plus 3pt 
\belowdisplayskip=0pt plus 3pt 
\belowdisplayshortskip=0pt plus 3pt 
\begin{displaymath}
\text{\texttt{a*b - c*d = (-d + a*b) - (-1 + c)*d}}
\end{displaymath}}}

If ideal membership cannot be verified, \texttt{certify} returns \texttt{False}.
This outcome can occur because of two reasons.
Either \texttt{claim} is simply not contained in the ideal generated by \texttt{assumptions}, 
or \texttt{certify}, which is an iterative procedure, had not been run for enough iterations to verify the ideal membership.
To avoid the latter situation, \texttt{certify} can be passed an optional argument \texttt{maxiter} to determine the maximal number of iterations it is run.
By default this value is set to $10$.

\begin{sageverbatim}
sage: assumptions = [a*b*a - a*b]
sage: claim = a*b^20*a - a*b^20
sage: certify(assumptions, claim)
Computing a (partial) Groebner basis and reducing the claims...

Starting iteration 5...
Starting iteration 10...
Failed! Not all ideal memberships could be verified.
\end{sageverbatim}
{{%
\abovedisplayskip=0pt plus 3pt 
\abovedisplayshortskip=0pt plus 3pt 
\belowdisplayskip=0pt plus 3pt 
\belowdisplayshortskip=0pt plus 3pt 
\begin{displaymath}
\text{\texttt{False}}
\end{displaymath}}}
\begin{sageverbatim}
sage: proof = certify(assumptions, claim, maxiter=20)
Computing a (partial) Groebner basis and reducing the claims...

Starting iteration 5...
Starting iteration 10...
Starting iteration 15...
Done! Ideal membership of all claims could be verified!
\end{sageverbatim}

\begin{remark}
Ideal membership in the free algebra is undecidable in general.
Thus, we can also not decide whether the number of iterations of \texttt{certify} was simply too low or whether \texttt{claim} is really not contained in the ideal.
\end{remark}

\subsection{Useful auxiliary functions for treating operator statements}

The package provides some auxiliary functions which help in constructing polynomials that commonly appear when treating operator statements.

\begin{itemize}
\item \texttt{pinv(a, b, a\char`_adj, b\char`_adj)}: generate the polynomials~\eqref{eq:moore-penrose-polynomials} encoding the four Penrose identities for \texttt{a} with Moore-Penrose inverse \texttt{b} and respective adjoints \texttt{a\char`_adj} and \texttt{b\char`_adj}.
\item \texttt{adj(f)}: compute the adjoint $\texttt{f}^*$ of a polynomial \texttt{f}. 
        Each variable \texttt{x} is replaced by \texttt{x\char`_adj}.
        Note that all variables \texttt{x} and \texttt{x\char`_adj} have to be defined as generators of the same \texttt{FreeAlgebra}.
\item \texttt{add\char`_adj(F)}: add to a list of polynomials \texttt{F} the corresponding adjoint elements.
\end{itemize}

\subsection{Quivers and detecting typos}

When encoding operator identities, the resulting polynomials can become quite intricate and it can easily happen that typos occur.
To detect typos, it can help to syntactically check if entered polynomials correspond to correctly translated operator identities, respecting the restrictions imposed by the domains and codomains.
To this end, the package allows to encode the domains and codomains in form of a directed labelled multigraph, called \emph{(labelled) quiver}.

\begin{figure}
\centering
\begin{tikzpicture}
 \matrix (m) [matrix of math nodes, column sep=2cm]
  {U & V & W\\};
 \path[->] (m-1-1) edge [bend left] node [auto] {$A$} (m-1-2);
 \path[->] (m-1-2) edge [bend left] node [auto] {$D$} (m-1-1);
 \path[->] (m-1-2) edge [bend left] node [auto] {$B$} (m-1-3);
 \path[->] (m-1-3) edge [bend left] node [auto] {$C$} (m-1-2);
\end{tikzpicture}
\caption{Quiver encoding domains and codomains of operators}
\label{fig:quiver}
\end{figure}
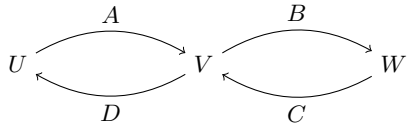
Computationally, a quiver is given by a list of triplets $(u,v,a)$, where $u$ and $v$ can be any symbols that encode the domain $U$ and the codomain $V$ of the basic operator $A$ and $a$ is the indeterminate representing $A$.
For example, a quiver encoding the situation of operators $A,B,C,D$ on spaces $U,V,W$ as in Figure~\ref{fig:quiver}, can be constructed as follows.

\begin{sageverbatim}
sage: F.<a,b,c,d> = FreeAlgebra(QQ)
sage: Q = Quiver([('U','V',a), ('V','W',b), ('W','V',c), ('V','U',d)])
sage: Q
\end{sageverbatim}
{{%
\abovedisplayskip=0pt plus 3pt 
\abovedisplayshortskip=0pt plus 3pt 
\begin{displaymath}
\text{\texttt{Labelled quiver with 3 vertices in the labels \{a, b, c, d\}}}
\end{displaymath}}}

One can easily check if a polynomial is compatible with the situation of operators encoded by a quiver.

\begin{sageverbatim}
sage: Q.is_compatible(a*b + c*d)
\end{sageverbatim}
{{%
\abovedisplayskip=0pt plus 3pt 
\abovedisplayshortskip=0pt plus 3pt 
\belowdisplayskip=0pt plus 3pt 
\belowdisplayshortskip=0pt plus 3pt 
\begin{displaymath}
\text{\texttt{False}}
\end{displaymath}}}
\begin{sageverbatim}
sage: Q.is_compatible(a*d + c*b)
\end{sageverbatim}
{{%
\abovedisplayskip=0pt plus 3pt 
\abovedisplayshortskip=0pt plus 3pt 
\begin{displaymath}
\text{\texttt{True}}
\end{displaymath}}}

A quiver can be handed as an optional argument to \texttt{certify}, which then checks all input polynomials for compatibility with the given quiver and raises an error if required.

\begin{sageverbatim}
sage: assumptions = [a*d, c*b]
# typo in the claim, c*b -> b*c
sage: claim = a*d - b*c
sage: certify(assumptions, claim, quiver=Q)
\end{sageverbatim}
{{%
\abovedisplayskip=0pt plus 3pt 
\abovedisplayshortskip=0pt plus 3pt 
\begin{displaymath}
\text{\texttt{ValueError: The claim a*d - b*c is not compatible with the quiver}}
\end{displaymath}}}

\subsection{Gröbner basis computations}
\label{appendix:ideal-membership}
Behind the scenes, the \texttt{certify} command computes Gröbner bases in the free algebra.
In this section, we present the methods of the package that allow to do such computations.

\subsubsection{Ideals and monomial orders}

The main data structure provided by the package is that of a (two-sided) ideal in the free algebra, called \texttt{NCIdeal}.
Such an ideal can be constructed from any finite set of noncommutative polynomials.

\begin{sageverbatim}
sage: F.<x,y,z> = FreeAlgebra(QQ)
sage: gens = [x*y*z - x*y, y*z*x*y - y]
sage: NCIdeal(gens)
\end{sageverbatim}
{{%
\abovedisplayskip=0pt plus 3pt 
\abovedisplayshortskip=0pt plus 3pt 
\begin{align*}
&\text{\texttt{NCIdeal (-x*y + x*y*z, -y + y*z*x*y) of Free Algebra on}}\\
&\text{\texttt{3 generators (x, y, z) over Rational Field with x < y < z}}
\end{align*}}}

Attached to an \texttt{NCIdeal} also comes a monomial order w.r.t.~which further computations are done.
By default, this is a degree left lexicographic order, where the indeterminates are sorted as in the parent \texttt{FreeAlgebra}.
The order of the variables can be individualised by providing a list as an optional argument \texttt{order}.
Furthermore, by providing a list of lists, block orders (also known as elimination orders) can be defined.
The order within each block is still degree left lexicographic and blocks are provided in ascending order.

\begin{sageverbatim}
sage: NCIdeal(gens, order=[y,x,z])
\end{sageverbatim}
{{%
\abovedisplayskip=0pt plus 3pt 
\abovedisplayshortskip=0pt plus 3pt 
\belowdisplayskip=0pt plus 3pt 
\belowdisplayshortskip=0pt plus 3pt 
\begin{align*}
&\text{\texttt{NCIdeal (-x*y + x*y*z, -y + y*z*x*y) of Free Algebra on}}\\
&\text{\texttt{3 generators (x, y, z) over Rational Field with y < x < z}}
\end{align*}}}
\begin{sageverbatim}
sage: NCIdeal(gens, order=[[y,x],[z]])
\end{sageverbatim}
{{%
\abovedisplayskip=0pt plus 3pt 
\abovedisplayshortskip=0pt plus 3pt 
\belowdisplayskip=0pt plus 3pt 
\belowdisplayshortskip=0pt plus 3pt 
\begin{align*}
&\text{\texttt{NCIdeal (-x*y + x*y*z, -y + y*z*x*y) of Free Algebra on}}\\
&\text{\texttt{3 generators (x, y, z) over Rational Field with y < x << z}}
\end{align*}}}
\subsubsection{Gröbner bases and normal forms}

For computing Gröbner bases, the class \texttt{NCIdeal} provides the method \texttt{groebner\char`_basis} with the following optional arguments:
\begin{itemize}
\item \texttt{maxiter} (default: 10): Maximal number of iterations executed.
\item \texttt{maxdeg} (default: $\infty$): Maximal degree of considered ambiguities.
\item \texttt{trace\char`_cofactors} (default: True): If cofactor representations of each Gröbner basis element in terms of the generators should be computed.
\item \texttt{criterion} (default: True): If Gebauer-Möller criteria~\cite{Xiu12} should be used to detect redundant ambiguities.
\item \texttt{reset} (default: True): If all internal data should be reset. If set to False, this allows to continue previous (partial) Gröbner basis computations.
\item \texttt{verbose} (default: 0): 'Verbosity' value determining the amount of information about the computational progress that is printed.
\end{itemize}

\begin{sageverbatim}
sage: F.<x,y> = FreeAlgebra(QQ)
sage: gens = [x*y*x - x*y, y*x*x*y - y]
sage: I = NCIdeal(gens)
sage: G = I.groebner_basis(); G
\end{sageverbatim}
{{%
\abovedisplayskip=0pt plus 3pt 
\abovedisplayshortskip=0pt plus 3pt 
\begin{align*}
&\text{\texttt{[- x*y + x*y*x, - y + y*x${}^2$*y, - y + y*x, - x*y + x*y${}^2$,}}\\
&\text{\texttt{- x*y + x*y${}^2$*x, - y + y${}^2$, - y + y${}^3$]}}
\end{align*}}}

We note that the polynomials output by the \texttt{groebner\char`_basis} routine are not \textsc{SageMath} noncommutative polynomials but our own
\texttt{NCPolynomial}s. 
They provide similar functionality as the native data structure (basic arithmetic, equality testing, coefficient/monomial extraction), but can additionally also store a cofactor representation.
In particular, the elements output by the \texttt{groebner\char`_basis} command all hold a cofactor representation w.r.t.~the generators of the \texttt{NCIdeal}.

\begin{sageverbatim}
sage: f = G[2]
sage: pretty_print_proof(f.cofactors(), I.gens())
\end{sageverbatim}
{{%
\abovedisplayskip=0pt plus 3pt 
\abovedisplayshortskip=0pt plus 3pt 
\belowdisplayskip=0pt minus 3pt 
\belowdisplayshortskip=0pt minus 3pt 
\begin{align*}
\text{\texttt{-y + y*x = y*x*(-x*y + x*y*x) + (-y + y*x${}^2$*y) - (-y + y*x${}^2$*y)*x}}
\end{align*}}}
\begin{remark}
To convert an \texttt{NCPolynomial} back into \textsc{SageMath}'s native data structure, our class provides the method \texttt{to\char`_native}.
Conversely, to convert a \textsc{SageMath} noncommutative polynomial \texttt{f} into an \texttt{NCPolynomial}, one can use \texttt{NCPolynomial(f)}.
\end{remark}

The package also allows to interreduce a set of \texttt{NCPolynomial}s using the command \texttt{interreduce}.
\begin{sageverbatim}
sage: interreduce(G)
\end{sageverbatim}
{{%
\abovedisplayskip=0pt plus 3pt 
\abovedisplayshortskip=0pt plus 3pt 
\belowdisplayskip=0pt plus 3pt 
\belowdisplayshortskip=0pt plus 3pt 
\begin{align*}
\text{\texttt{[- y + y*x, - y + y${}^2$]}}
\end{align*}}}

To compute the normal form of an element \texttt{f} w.r.t.~the generators of an \texttt{NCIdeal}, the class provides the method \texttt{reduced\char`_form}.
The output of this method is an \texttt{NCPolynomial} \texttt{g} holding a cofactor representation of the difference \texttt{f - g} w.r.t.~the generators of the \texttt{NCIdeal} 
The method \texttt{reduced\char`_form} accepts the same optional arguments as \texttt{groebner\char`_basis}.

\begin{sageverbatim}
sage: f = I.reduced_form(y^2 - y); f
\end{sageverbatim}
{{%
\abovedisplayskip=0pt plus 3pt 
\abovedisplayshortskip=0pt plus 3pt 
\belowdisplayskip=0pt plus 3pt 
\belowdisplayshortskip=0pt plus 3pt 
\begin{align*}
\text{\texttt{0}}
\end{align*}}}
\begin{sageverbatim}
sage: pretty_print_proof(f.cofactors(), I.gens())
\end{sageverbatim}
{{%
\abovedisplayskip=0pt plus 3pt 
\abovedisplayshortskip=0pt plus 3pt 
\belowdisplayskip=0pt plus 3pt 
\belowdisplayshortskip=0pt plus 3pt 
\begin{align*}
\text{\texttt{-y + y${}^2${ }={ }}} 
&\text{\texttt{(-y + y*x${}^2$*y) - y*x*(-x*y + x*y*x)*y - (-y + y*x${}^2$*y)*y}}\\
&\text{\texttt{- y*x*(-x*y + x*y*x)*x*y + (-y + y*x${}^2$*y)*x${}^2$*y}}
\end{align*}}}
\begin{sageverbatim}
sage: I.reduced_form(y^2)
\end{sageverbatim}
{{%
\abovedisplayskip=0pt minus 5pt 
\abovedisplayshortskip=0pt minus 5pt 
\belowdisplayskip=0pt minus 5pt
\belowdisplayshortskip=0pt minus 5pt
\begin{align*}
\text{\texttt{y}}
\end{align*}}}
\subsection{Heuristics for finding polynomials of certain form}
\label{appendix:heuristics}

One of the main functionalities provided by the package are dedicated heuristics for systematically searching for polynomials of certain form in an \texttt{NCIdeal}.
To this end, the class \texttt{NCIdeal} provides the method \texttt{find\char`_equivalent\char`_expression(f)}, which searches for elements of the form \texttt{f - g} with arbitrary \texttt{g} in an \texttt{NCIdeal}. 
It accepts the following optional arguments:
\begin{itemize}
\item All optional arguments that also \texttt{groebner\char`_basis} accepts with the same effects.
\item \texttt{order}: A monomial order w.r.t.~which the computation is executed. The argument has to be provided like a custom order when defining an \texttt{NCIdeal} (see Sec.~\ref{appendix:ideal-membership}).
\item \texttt{heuristic} (default: 'groebner'): Determines the heuristic used. Available are 
  \begin{itemize}
    \item 'naive': Try exhaustively all monomials \texttt{m} up to a degree bound and check if \texttt{f - m} is in the ideal.
    \item 'groebner': Enumerate a Gröbner basis and search in the Gröbner basis for suitable elements containing \texttt{f}.
    \item 'subalgebra': Intersect the two-sided ideal with a subalgebra to find suitable elements.
    \item 'right-ideal'/'left-ideal': Intersect the two-sided ideal with a right/left ideal to find suitable elements.
   \end{itemize}
\item \texttt{prefix} (default: None): A term \texttt{p} providing the prefix of \texttt{g}, i.e., the heuristic looks for elements of the form \texttt{f - p*h} with arbitrary \texttt{h} (required for heuristic 'right-ideal').
\item \texttt{suffix} (default: None): A term \texttt{s} providing the suffix of \texttt{g}, i.e., the heuristic looks for elements of the form \texttt{f - h*s} with arbitrary \texttt{h} (required for heuristic 'left-ideal').
\item \texttt{degbound} (default: 5): Some heuristics only compute up to a fixed degree bound. This argument allows to change this degree bound.
\item \texttt{quiver} (default: None): Use a quiver to restrict the search space only to polynomials that are compatible with this quiver.
\end{itemize}

\begin{sageverbatim}
sage: F.<a,b,c,d> = FreeAlgebra(QQ)
sage: gens = [a*b*a-a, b*a*b-b, a*b-c*d, b*a-d*c, c*d*c-c, d*c*d-d]
sage: I = NCIdeal(gens)
sage: I.find_equivalent_expression(a*b)
\end{sageverbatim}
{{%
\abovedisplayskip=0pt plus 3pt 
\abovedisplayshortskip=0pt plus 3pt 
\belowdisplayskip=0pt plus 3pt 
\belowdisplayshortskip=0pt plus 3pt 
\begin{align*}
\text{\texttt{[-a*b + c*d]}}
\end{align*}}}
\begin{sageverbatim}
sage: I.find_equivalent_expression(a*b, heuristic='naive', suffix=b)
\end{sageverbatim}
{{%
\abovedisplayskip=0pt plus 3pt 
\abovedisplayshortskip=0pt plus 3pt 
\belowdisplayskip=0pt plus 3pt 
\belowdisplayshortskip=0pt plus 3pt 
\begin{align*}
\text{\texttt{[a*b - c*d*a*b]}}
\end{align*}}}
\begin{sageverbatim}
sage: I.find_equivalent_expression(a*b, heuristic='right-ideal', 
....: prefix=a*b)
\end{sageverbatim}
{{%
\abovedisplayskip=0pt plus 3pt 
\abovedisplayshortskip=0pt plus 3pt 
\begin{align*}
\text{\texttt{[- a*b + a*b*c*d, - a*b + a*b*a*b]}}
\end{align*}}}
Additionally, the class \texttt{NCIdeal} provides methods for applying cancellability.
\begin{itemize}
\item \texttt{I.apply\char`_left\char`_cancellability(a, b)}: Search for elements of the form \texttt{a*b*f} in \texttt{I} and return \texttt{b*f}.
\item \texttt{I.apply\char`_right\char`_cancellability(a, b)}: Search for elements of the form \texttt{f*a*b} in \texttt{I} and return \texttt{f*a}.
\end{itemize}
Both methods can be given an optional argument \texttt{heuristic} to determine the used search heuristic. Available are 'subalgebra', 'one-sided', and 'two-sided' (default: 'subalgebra').

\begin{sageverbatim}
sage: I.apply_left_cancellability(c, a)
\end{sageverbatim}
{{%
\abovedisplayskip=0pt plus 3pt 
\abovedisplayshortskip=0pt plus 3pt 
\belowdisplayskip=0pt plus 3pt 
\belowdisplayshortskip=0pt plus 3pt 
\begin{align*}
\text{\texttt{[- a + a*b*a, - a${}^2$ + a*d*c*a]}}
\end{align*}}}
\begin{sageverbatim}
#verify ideal membership to check correctness of result
sage: I.reduced_form(c*(-a^2 + a*d*c*a))
\end{sageverbatim}
{{%
\abovedisplayskip=0pt plus 3pt 
\abovedisplayshortskip=0pt plus 3pt 
\belowdisplayskip=0pt plus 3pt 
\belowdisplayshortskip=0pt plus 3pt 
\begin{align*}
\text{\texttt{0}}
\end{align*}}}
\begin{sageverbatim}
sage: I.apply_right_cancellability(a*b, d*a, heuristic='two-sided',
....: maxiter=5)
\end{sageverbatim}
{{%
\abovedisplayskip=0pt plus 3pt 
\abovedisplayshortskip=0pt plus 3pt 
\belowdisplayskip=0pt plus 3pt 
\belowdisplayshortskip=0pt plus 3pt 
\begin{align*}
\text{\texttt{[- a*b + a*b*a*b, - a*b + c*d*a*b]}}
\end{align*}}}
\begin{sageverbatim}
#verify ideal membership to check correctness of result
sage: I.reduced_form((-a*b + c*d*a*b)*c*d)
\end{sageverbatim}
{{%
\abovedisplayskip=0pt plus 3pt 
\abovedisplayshortskip=0pt plus 3pt 
\belowdisplayskip=0pt plus 3pt 
\belowdisplayshortskip=0pt plus 3pt 
\begin{align*}
\text{\texttt{0}}
\end{align*}}}

\end{document}